\documentclass[prl,twocolumn,showpacs,floatfix,superscriptaddress]{revtex4}

\newcommand{\be}{\begin{equation}}
\newcommand{\ee}{\end{equation}}

\newcommand{\down}{\mathopen|\downarrow\rangle}
\newcommand{\up}{\mathopen|\uparrow\rangle}
\newcommand{\ij}{i\kern -0.08em j}
\newcommand{\po}{\phantom{1}}

\usepackage{graphicx,amsmath,amssymb}
\usepackage{bm}

\begin{document}

\title{Four-qubit device with mixed couplings}

\author{M.~Grajcar}
\affiliation{%
Institute for Physical High Technology, P.O. Box 100239, D-07702
Jena, Germany}
\affiliation{%
Department of Solid State Physics,
Comenius University, SK-84248 Bratislava, Slovakia}
\affiliation{%
Center of Excellence of the Slovak Academy of Sciences (CENG), Slovakia}

\author{A.~Izmalkov}
\affiliation{%
Institute for Physical High Technology, P.O. Box 100239, D-07702
Jena, Germany}

\author{S.~H.~W. van der Ploeg}
\affiliation{%
Institute for Physical High Technology, P.O. Box 100239, D-07702
Jena, Germany}
\affiliation{MESA+ Research Institute and Faculty of
Science and Technology, University of Twente, P.O. Box 217, 7500
AE Enschede, The Netherlands}

\author{S.~Linzen}
\affiliation{%
Institute for Physical High Technology, P.O. Box 100239, D-07702
Jena, Germany}

\author{T.~Plecenik}
\affiliation{%
Institute for Physical High Technology, P.O. Box 100239, D-07702
Jena, Germany}
\affiliation {%
Department of Solid State Physics,
Comenius University, SK-84248 Bratislava, Slovakia}

\author{Th.~Wagner}
\affiliation{%
Institute for Physical High Technology, P.O. Box 100239, D-07702
Jena, Germany}

\author {U.~H\"ubner}
\affiliation{%
Institute for Physical High Technology, P.O. Box 100239, D-07702
Jena, Germany}

\author{E.~Il'ichev}
\email{ilichev@ipht-jena.de}
\affiliation{%
Institute for Physical High Technology, P.O. Box 100239, D-07702
Jena, Germany}

\author{H.-G. Meyer}
\affiliation{%
Institute for Physical High Technology, P.O. Box 100239, D-07702
Jena, Germany}

\author{A.~Yu. Smirnov}
\affiliation{D-Wave Systems Inc., 100-4401 Still Creek Drive, Burnaby, B.C., V5C 6G9 Canada}

\author{Peter J. Love}
\affiliation{D-Wave Systems Inc., 100-4401 Still Creek Drive, Burnaby, B.C., V5C 6G9 Canada}

\author{Alec~Maassen van den Brink}
\email{alec@dwavesys.com}
\affiliation{D-Wave Systems Inc., 100-4401 Still Creek Drive, Burnaby, B.C., V5C 6G9 Canada}

\author{M.~H.~S. Amin}
\affiliation{D-Wave Systems Inc., 100-4401 Still Creek Drive, Burnaby, B.C., V5C 6G9 Canada}

\author{S.~Uchaikin}
\affiliation{D-Wave Systems Inc., 100-4401 Still Creek Drive, Burnaby, B.C., V5C 6G9 Canada}

\author{A.~M. Zagoskin}
\email{zagoskin@physics.ubc.ca}
\affiliation{Physics and Astronomy Dept., The University of British Columbia, 6224 Agricultural Rd., Vancouver, B.C., V6T 1Z1 Canada}

\date{\today}

\begin{abstract}
We present the first experimental results on a device with more than two superconducting qubits. The circuit consists of four three-junction flux qubits, with simultaneous ferro- and antiferromagnetic coupling implemented using shared Josephson junctions. Its response, which is dominated by the ground state, is characterized using low-frequency impedance measurement with a superconducting tank circuit coupled to the qubits. The results are found to be in excellent agreement with the quantum-mechanical predictions.
\end{abstract}

\pacs{85.25.Dq, 85.25.Cp, 03.67.Lx}

\maketitle

\begin{figure}[t]
\includegraphics[width=7cm]{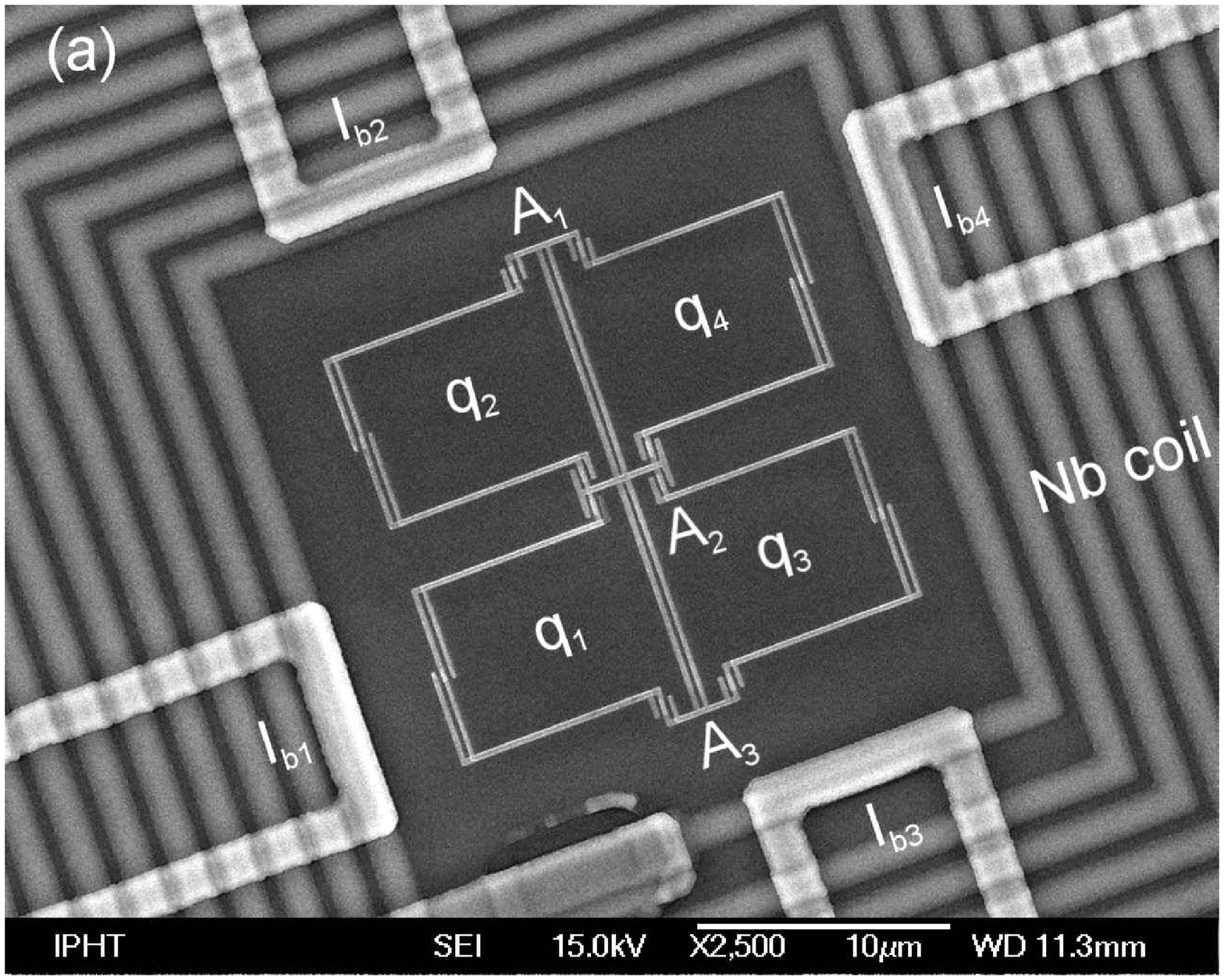}

\vspace{3mm}
\includegraphics[width=5cm]{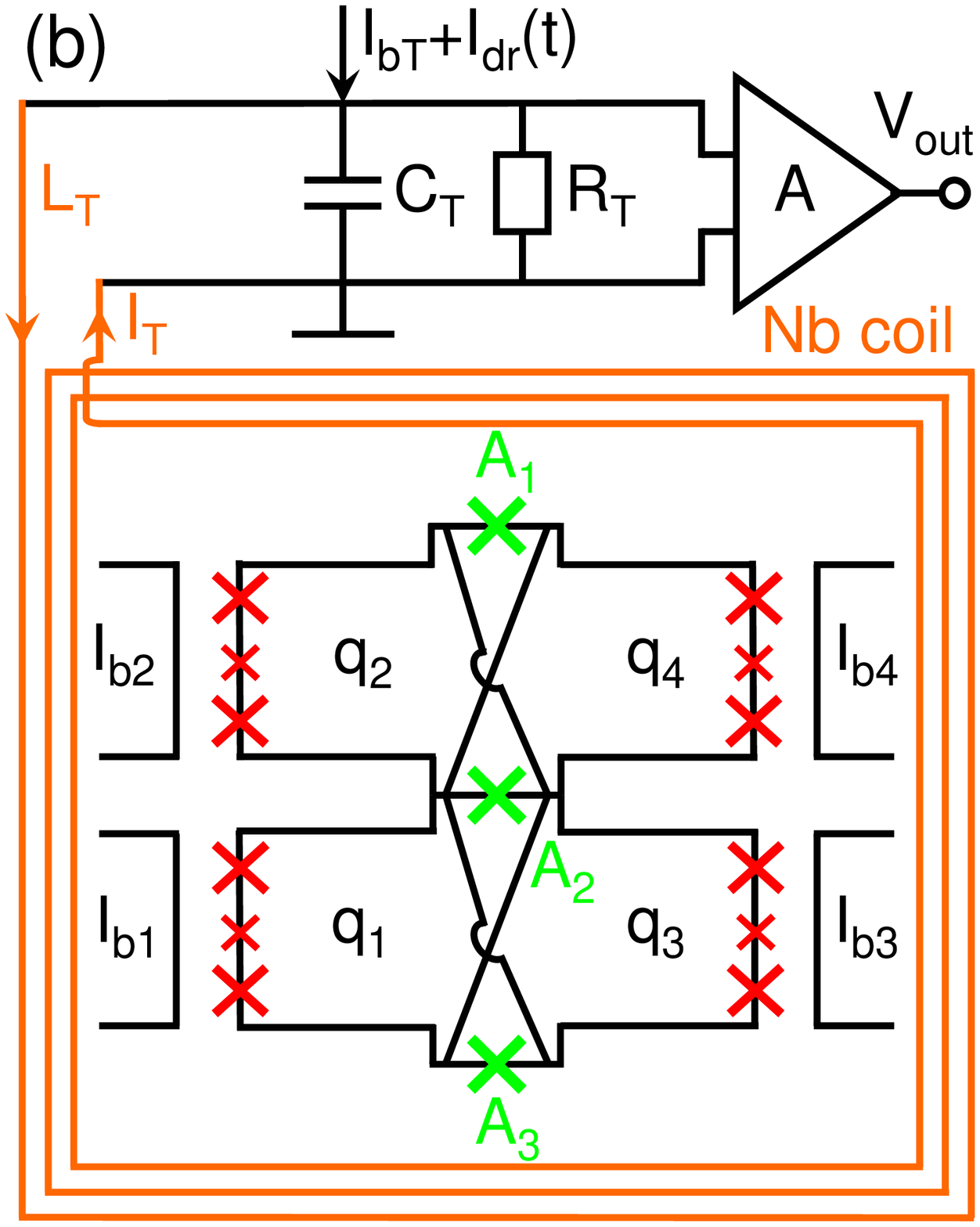}
\caption{(a) Electron micrograph of the sample. The central junctions $A_1$--$A_3$ couple the Al qubits $q_1$--$q_4$. The surrounding Nb coil is part of the $LC$ tank circuit, used for both measurement and global flux biasing. The Nb lines $I_\text{b1--b4}$ allow asymmetric bias tuning. (b)~Schematic diagram. The driving current $I_\mathrm{dr}(t)$ is much smaller than the oscillations in $I_\mathrm{T}$ and only needs to compensate for the latter's losses.}
\label{fig1}
\end{figure}

The implementation of one- and two-qubit gates in superconducting qubit prototypes~\cite{Nakamura1999} has confirmed their utility for quantum computation. Such qubits are readily fabricated, and highly scalable in principle. A prominent subcategory consists of superconducting loops interrupted by three Josephson junctions, so-called 3JJ flux qubits~\cite{Mooij1999}. If threaded by a near-degenerate magnetic bias flux $\Phi^\mathrm{x}\approx\frac{1}{2}\Phi_0$ ($\Phi_0$ is the flux quantum), such devices have quantum states which are superpositions of clockwise ($\down$) and counterclockwise ($\up$) circulating supercurrent~$I_\mathrm{p}$. The small area of 3JJ qubits reduces their coupling to environmental magnetic noise, making comparatively long coherence times possible~\cite{coherence}, but also limits the strength of their inductive coupling \cite{Mooij1999,Izmalkov2004a}. This can be overcome using direct galvanic coupling through a shared Josephson junction~\cite{Levitov2001}. When reporting its experimental realization~\cite{Grajcar2005a}, we mentioned that direct coupling can have additional advantages. First, the coupling strength can be varied independently of the sample geometry by changing the shared junction's critical current. Second, while direct coupling normally has the antiferromagnetic (AF) sign just as in the inductive case, a ``twisted" design (joining two qubit loops in a ``$\infty$'' shape with crossing leads in the center, see Fig.~\ref{fig1}b) features ferromagnetic (FM) coupling~\cite{YNN}.

In this Letter, we pursue this by studying a \emph{four}-qubit circuit in which the two types of coupling co-exist. This is very promising from the perspectives of realizing nontrivial Ising-spin systems~\cite{barahona} and scalable adiabatic quantum computing (AQC)~\cite{Farhi2000,Grajcar2005}. However, while our circuit behaves in full agreement with quantum mechanics and features excellent multi-qubit bias control, the measurements presented are essentially equilibrium. The system's effective low-energy (pseudospin) Hamiltonian is
\be
 H= -\sum_{i=1}^4 [\epsilon_i \sigma_z^{(i)} + \Delta_i
 \sigma_x^{(i)}] + \sum_{1\le i<j\le4} J_{ij}
 \sigma_z^{(i)}\sigma_z^{(j)}\;,\label{H}
\ee
where $\epsilon_i$ is the bias on qubit~$i$ (implemented through a flux bias $\Phi^\mathrm{x}{-}\frac{1}{2}\Phi_0$), $\Delta_i$ is its tunnelling amplitude, $J_{ij}$ is the coupling energy between qubits $i$ and~$j$, and $\sigma_z$, $\sigma_x$ are Pauli matrices in the span of $\down$ ($\sigma_z=-1$) and $\up$ ($\sigma_z=1$). Equation~(\ref{H}) adequately describes a system of flux qubits \cite{Mooij1999,Izmalkov2004a,Chiorescu2003}, although possible limitations~\cite{MaassenvandenBrink2005b} remain to be investigated.% \textbf{cite NJP when ready}

Our approach follows earlier work using the impedance measurement technique (IMT)~\cite{Greenberg2002}. This method measures impedance shifts in a high-quality superconducting $LC$ (tank) circuit inductively coupled to the system. For one qubit~\cite{Grajcar2004}, the average loop current $I=(|b|^2{-}|a|^2)I_\mathrm{p}$ in an eigenstate $a\down+b\up$ changes direction at the anticrossing point $\Phi^\mathrm{x}=\frac{1}{2}\Phi_0$ (where $|a|=|b|$, i.e., the pseudospin ground state flips from $\down$ to $\up$ and vice versa for the excited state). This causes a peak in the qubit susceptibility $\chi=dI/d\Phi^\mathrm{x}$; since $dI/d\Phi^\mathrm{x} \propto d^2E/d(\Phi^\mathrm{x})^2$, where $E$ is the average energy and $d_{\Phi^\mathrm{x}}$ is adiabatic (note~15 in Ref.~\cite{Grajcar2004}), this also follows from the large energy-band curvature at the anticrossing. Note that the very presence of such peaks is already a quantum effect, since the classical $\chi(\Phi^\mathrm{x})$ curves feature hysteretic loops~\cite{Grajcar2005} in the absence of flux tunneling. Due to the qubit--tank coupling, $\chi$ affects the tank's eigenfrequency $\omega_\mathrm{T}$ and hence, under resonant driving, its current--voltage phase angle~$\Theta\propto\chi$. Thus, the width and height of such an \emph{IMT-peak} in~$\Theta$ provide information about the anticrossing, i.e., about $\Delta$ and~$I_\mathrm{p}$.

For multiple qubits, a more general analysis yields~\cite{Smirnov2003a}
\be
  \tan\Theta = -2\frac{Q_\mathrm{T}}{L_\mathrm{T}}\sum_{\mu<\nu}
  \frac{\rho_{\mu} - \rho_{\nu}}{ E_{\nu} - E_{\mu} } R_{\mu \nu}\;,
  \label{tantheta}
\ee
where $L_\mathrm{T}$ and $Q_\mathrm{T}$ are the tank's inductance and quality factor respectively, $\rho_{\mu} = e^{-E_{\mu}/T}/\sum_{\nu}e^{-E_{\nu}/T}$ is the Boltzmann factor of the eigenstate $|\mu\rangle$ of $H$ in (\ref{H}), and with the matrix element
\begin{eqnarray}
  R_{\mu\nu} = \sum_{i,j=1}^4 \lambda_i \lambda_j\, \langle \mu
  |\sigma_{z}^{(i)}|\nu \rangle\, \langle \nu |\sigma_{z}^{(j)}|\mu
  \rangle\;; \label{R}
\end{eqnarray}
$\lambda_i$ are tank--qubit coupling coefficients in $H_{\rm int}=I_\mathrm{rf} \sum_i \lambda_i\sigma_z^{(i)}$, and $I_\mathrm{rf}$ is the oscillating component of the tank current $I_\mathrm{T}=I_\mathrm{bT}+I_\mathrm{rf}(t)$. The term $i=j$ produces the IMT-peak corresponding to qubit~$i$. Cross terms $i\ne j$, on the other hand, appear only if $|\mu\rangle$ or $|\nu\rangle$ has nonzero entanglement between qubits $i$ and~$j$ due to their interaction. For two weakly coupled qubits~\cite{Izmalkov2004a}, the effect of interaction can be understood as a correction. The cross terms in $R_{\mu\nu}$ then change the shape of overlapping peaks at co-degeneracies [i.e., when the $\epsilon_i$ in (\ref{H}) are chosen such that both qubits flip simultaneously], which turn out to be smaller (larger) than the sum of the two individual peaks for AF (FM) coupling~\cite{ex-def}.

\begin{figure}[t]
\includegraphics[width=8cm]{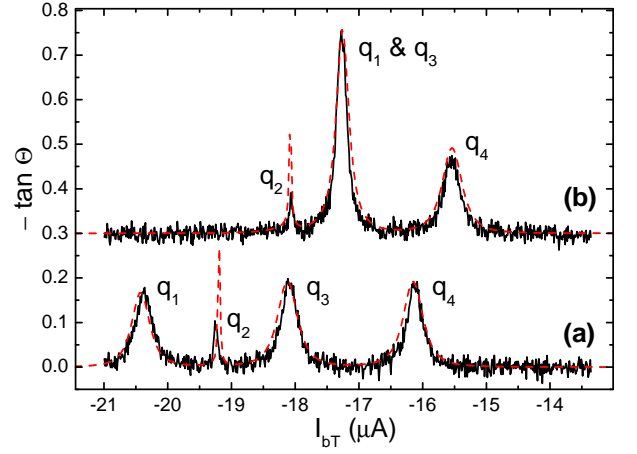}
\caption{(a) $-\tan \Theta(I_\mathrm{bT})$ at $I_\mathrm{b1}=700~\mu$A, $I_\mathrm{b2}=10~\mu$A, and $I_\mathrm{b4}=250~\mu$A ($I_\mathrm{b3}=0$ throughout). The four qubit peaks do not overlap. (b)~The same graph (vertically shifted for clarity) but at $I_\mathrm{b1}=268~\mu$A, where the peaks corresponding to $q_1$ and $q_3$ overlap. The height increase compared to the sum of the two non-overlapping peaks is a signature of FM coupling. Dashed lines are theoretical fits. The fit for $q_2$ is poor because the relatively large tank amplitude washes out the experimental peak~\protect\cite{Greenberg2002}; a better fit has been obtained for smaller amplitudes (not shown).}
\label{fig2}
\end{figure}

Here, the analysis needs to be extended both beyond two qubits and to stronger coupling (particularly $J>\nobreak\Delta$), which qualitatively changes the locus of the IMT-peaks, as can already be understood classically: neglecting the $\Delta_i$ in (\ref{H}), the degeneracy points of the classical multi-qubit flux states are seen to depend on the~$J_{ij}$~\cite{Grajcar2005a}. One then proceeds by fitting the coefficients of $H$ in~(\ref{H}), see below.

Figure~\ref{fig1}a shows the sample with four Al qubit loops inside a Nb pancake coil. The areas of the Al loops are $10\times7.5~\mu$m$^2$. Two junctions in each qubit are typically $650\times150$~nm$^2$, while the third is $\sim$25\% smaller~\cite{Mooij1999}. The Nb coil has $L_\mathrm{T}=105$~nH, and together with an external capacitance $C_\mathrm{T}=470$~pF forms a parallel tank circuit with $\omega_\mathrm{T}/2\pi=22.703$~MHz and $Q_\mathrm{T}=\omega_\mathrm{T} R_\mathrm{T} C_\mathrm{T}=400$, where $R_\mathrm{T}$ is the effective resistance. The Al loops were fabricated by $e$\nobreakdash-beam lithography and conventional shadow evaporation, and the Nb coil by $e$-beam lithography and CF$_4$ reactive-ion etching.

The qubits are coupled to each other both magnetically and via a shared junction, consisting of parallel junctions $A_1$--$A_3$  (Fig.~\ref{fig1}). For a coupling junction with large critical current~$I_\mathrm{c}$, and assuming identical qubits, $J=\hbar I_\mathrm{p}^2/2eI_\mathrm{c}$~\cite{Grajcar2005a}. For our system, $I_\mathrm{p}\sim 250$~nA. Due to the relative twist~\cite{Grajcar2005a} between the qubit loops (Fig.~1b), the couplings $q_1$--$q_3$ and $q_2$--$q_4$ are FM ($J_{ij}<\nobreak0$), while the others are AF ($J_{ij}>0$). The biases $\epsilon_i$ can be varied by adjusting the currents $I_\text{b1--b4}$ through the Nb wires, as well as the dc current $I_\mathrm{bT}$ through the tank coil. The qubit--coil mutual inductances follow from the flux periodicity as $M_{q1,\mathrm{T}}\approx M_{q3,\mathrm{T}}\approx38.2$~pH, $M_{q2,\mathrm{T}}=38.6$~pH, and $M_{q4,\mathrm{T}}=38.9$~pH.

\begin{figure}[t]
\includegraphics[width=7cm]{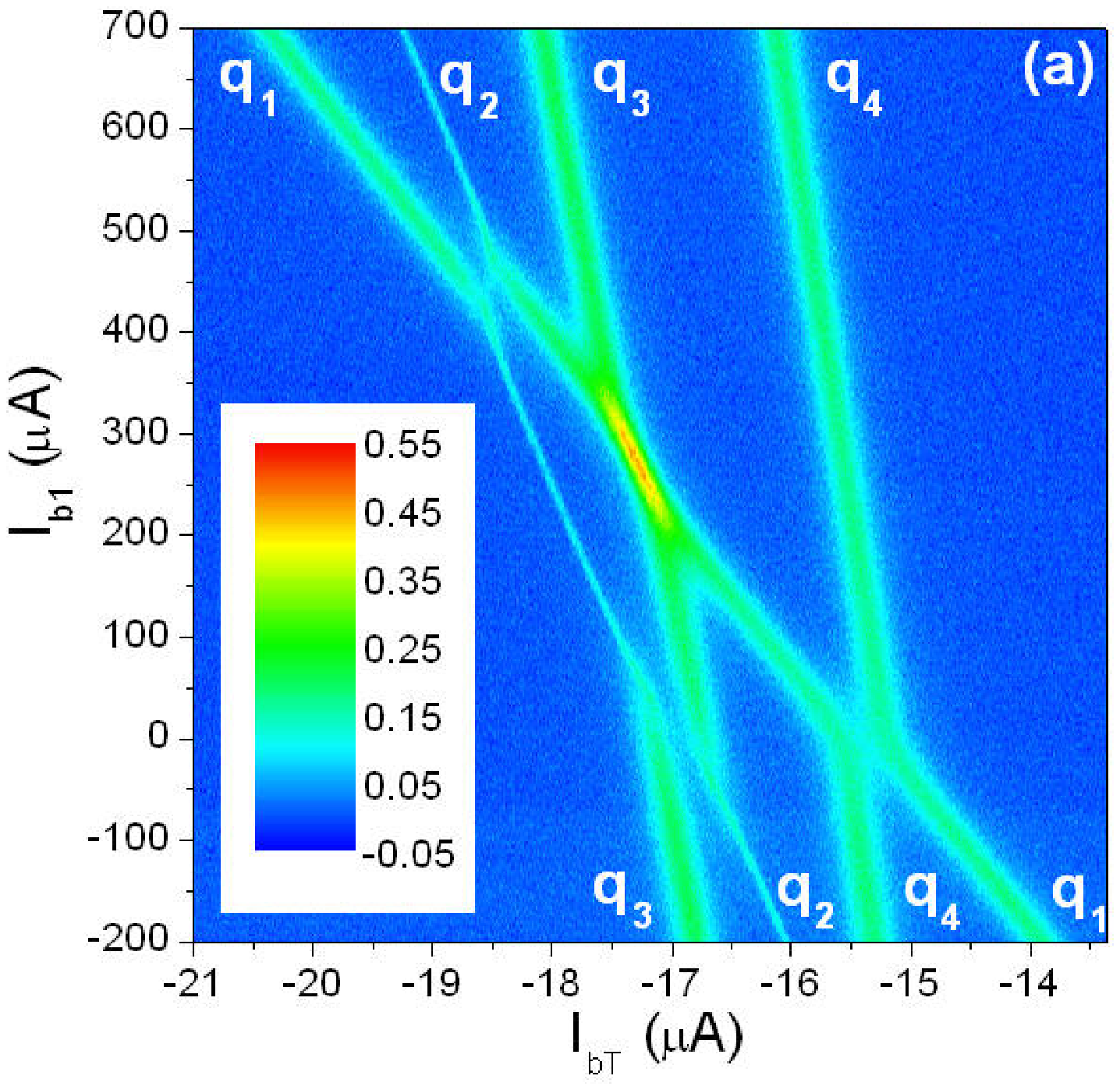}

\vspace{2mm}
\includegraphics[width=7cm]{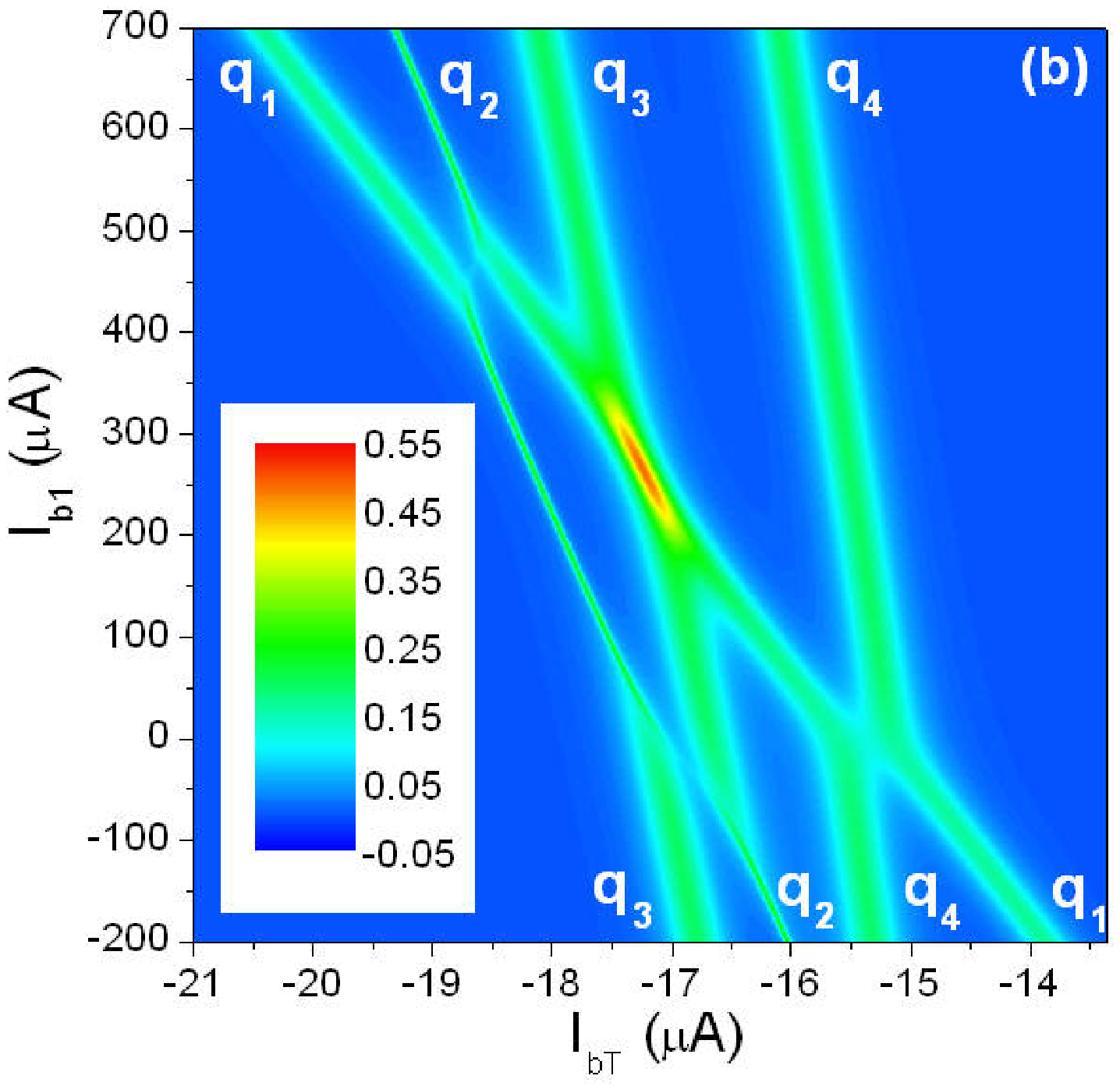}
\caption{(a) $-\tan \Theta(I_\mathrm{bT},I_\mathrm{b1})$ at $I_\mathrm{b2}=10~\mu$A and $I_\mathrm{b4}=250~\mu$A. The repulsion between the traces ($q_1,q_2$), ($q_2,q_3$), and ($q_1,q_4$) shows AF coupling between the qubits in those pairs, while the merging of traces ($q_1,q_3$) demonstrates FM coupling. The line widths are proportional to the tunnelling amplitudes of the individual qubits. Since the qubits' sensitivity to $I_\mathrm{b1}$ decreases in the order $q_1$, $q_2$, $q_3$, $q_4$ (cf.\ Fig.~\ref{fig1}a), the various slopes allow one to identify the trace belonging to each qubit. (b)~Theoretical graph for the same parameters.}
\label{fig3}
\end{figure}

\begin{figure}
\includegraphics[width=7cm]{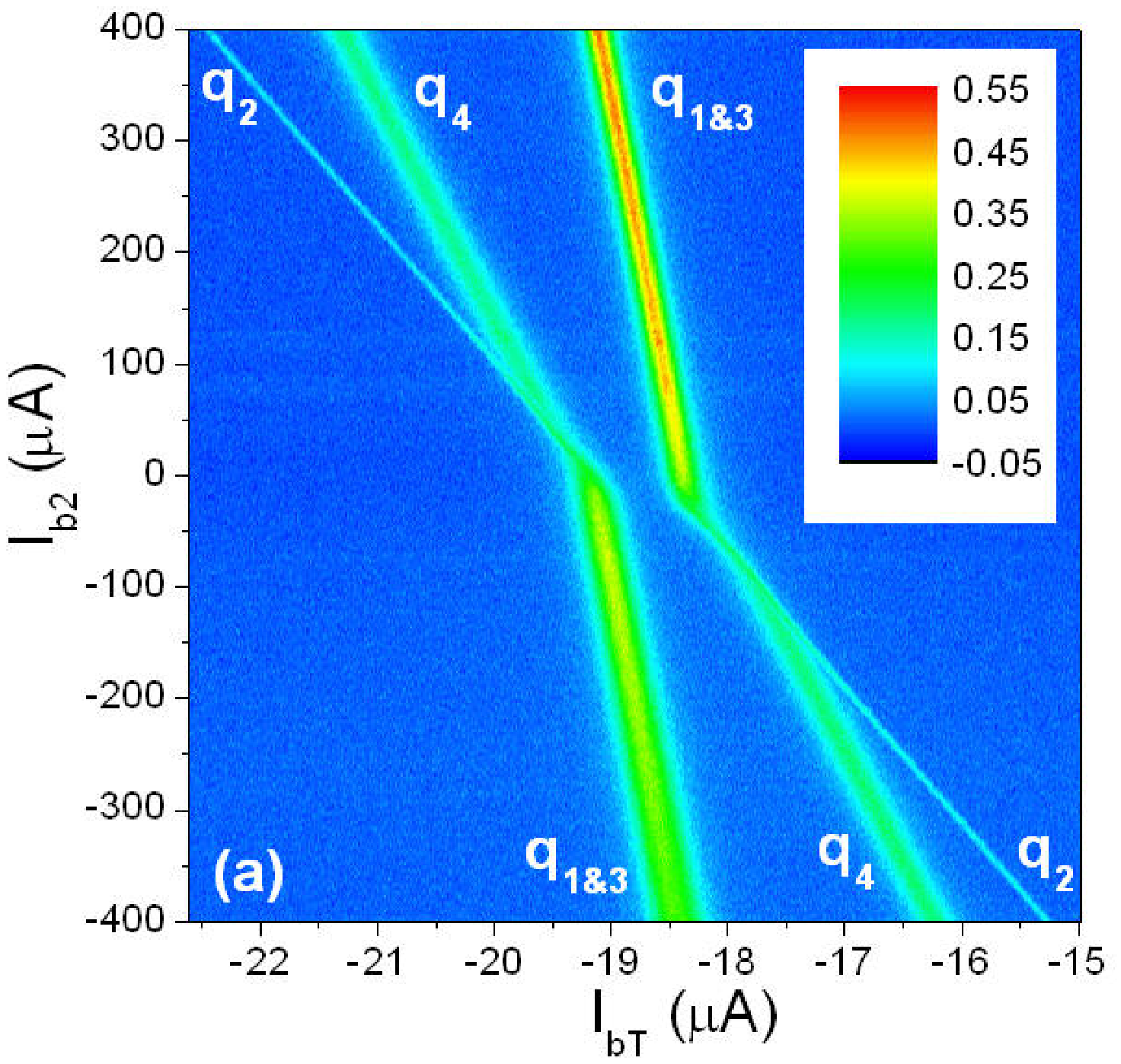}

\vspace{2mm}
\includegraphics[width=7cm]{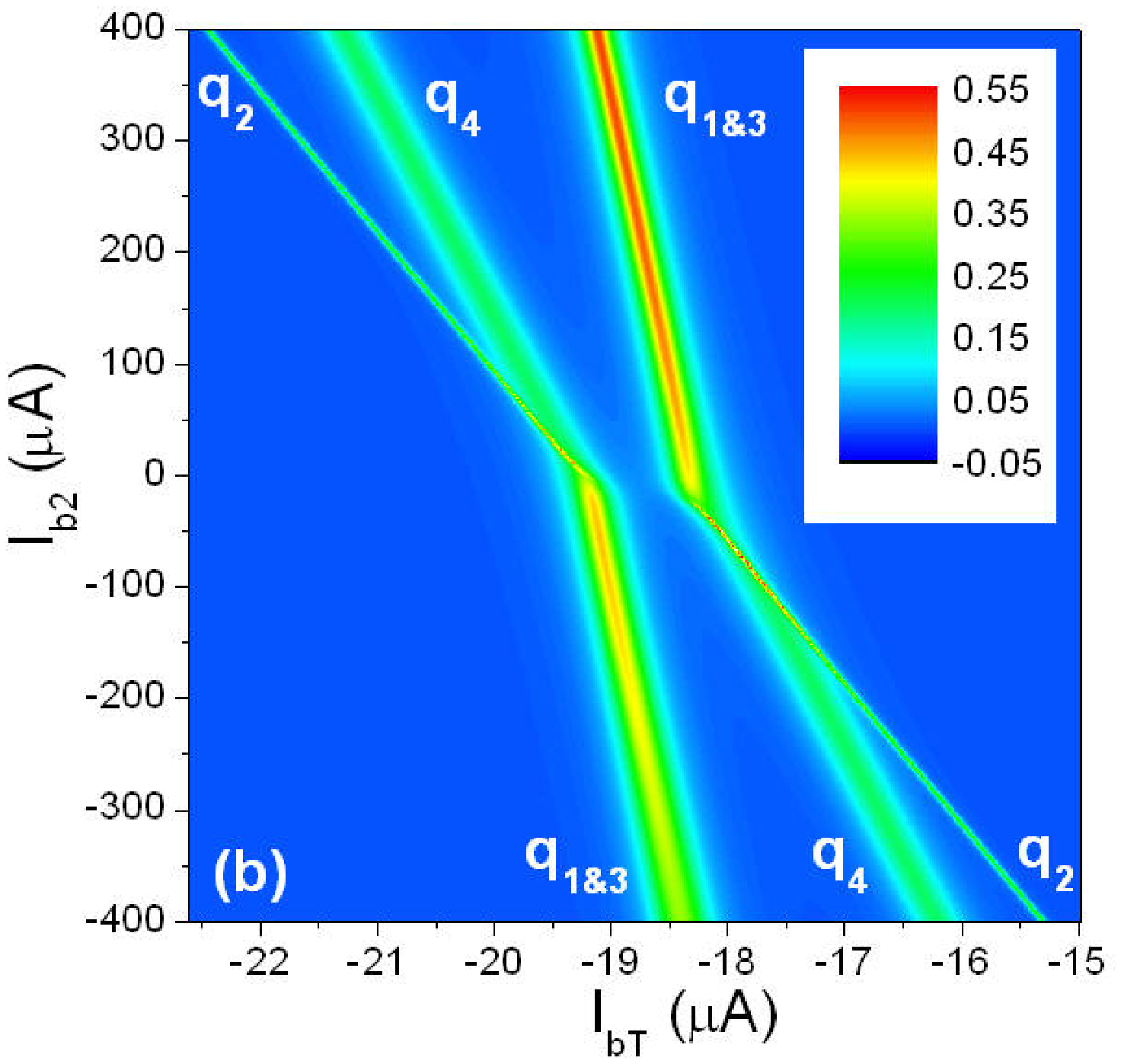}
\caption{Same as in Fig.~\ref{fig3}, but as a function of $I_\mathrm{bT}$ and $I_\mathrm{b2}$, at $I_\mathrm{b1}=-400~\mu$A and $I_\mathrm{b4}=-50~\mu$A. The skipped crossing at the center corresponds to four-qubit co-degeneracy. Due to their FM coupling, the $q_1$- and $q_3$-traces merge over an extended region. The qubits' sensitivity to $I_\mathrm{b2}$ decreases in the order $q_2$, $q_4$, $q_{1,3}$.}
\label{fig4}
\end{figure}

The sample was thermally anchored to the mixing chamber of a dilution refrigerator at a temperature $T_\mathrm{mix}\approx10$~mK. In general the effective $T$ is affected by noise on external leads and so is larger than~$T_\mathrm{mix}$. Often, $T_\mathrm{mix}$ alone is given, since $T$ is difficult to determine. One of the advantages of IMT is that it allows one to determine effective temperatures~\cite{Grajcar2005a,Grajcar2004}. From the best theoretical fit we estimate $T\approx70$~mK, well within the expected range.

Figure~\ref{fig2} shows peaks in $-\tan\Theta$ as the overall flux is scanned by changing~$I_\mathrm{bT}$. For most biases, four peaks can be distinguished (Fig.~\ref{fig2}a), each corresponding to degeneracy in one of the qubits while the others provide a semiclassical static background field. The narrow peak for $q_2$ indicates a small $\Delta_2$ ($\sim$12~mK), and incidentally confirms that the width of the other traces is not resolution-limited. When $q_1$ and $q_3$ are co-degenerate (Fig.~\ref{fig2}b), the resulting peak exceeds the sum of their individual peaks, indicating strong FM coupling as mentioned above (this is also true for $q_2$ and~$q_4$).

Figure \ref{fig3}a plots $-\tan\Theta(I_\mathrm{bT},I_\mathrm{b1})$. The four traces are IMT-peaks, each corresponding to a single-qubit anticrossing. Since $I_\mathrm{bT}$ biases all qubits almost equally, the trace with the greatest sensitivity to $I_\mathrm{b1}$ can be ascribed to $q_1$ etc. Combining several such cross-sections of the total IMT response, one reconstructs the biasing coefficients in $\Phi_i^\mathrm{x}=M_{qi,\mathrm{T}}I_\mathrm{bT}+\sum_j M'_{qi,\mathrm{b}j}I_{\mathrm{b}j}$. The centers of the traces mark the boundaries of the \emph{classical} stability diagram. Each region between the traces corresponds to a different (minimum-potential) flux state of the qubits~\cite{complete}. The shifts of the traces after each (skipped) crossing depend on the~$J_{ij}$. The peak shapes, on the other hand, carry information about the~$\Delta_i$. As expected, the traces corresponding to two qubits with AF (FM) coupling repel (attract) each other, and in the FM case merge over a certain distance. Fitting the data to Eq.~(\ref{H}) yields $\Delta_1=147$, $\Delta_2=12$, $\Delta_3=163$, $\Delta_4=165$, and $J_{12}=J_{34}=163$, $J_{14}=J_{23}=155$, $J_{13}=J_{24}=-62$ (all in mK), and
\be
  M'=\begin{pmatrix} 247 & \po71 & \cdot & \po68 \\ 143 & 309 & \cdot & \po84 \\
                     \po70 & \po87 & \cdot & 182 \\ \po50 & 195 & \cdot & 355
     \end{pmatrix}\mathrm{fH}\;,
\ee
where the missing entries correspond to $I_\mathrm{b3}$ which was never used. The theory (Fig.~\ref{fig3}b) accurately describes the system's behavior over the whole parameter space, especially at the co-degeneracies.

Figure \ref{fig4}a shows $-\tan\Theta(I_\mathrm{bT},I_\mathrm{b2})$ for a case when all four qubits become degenerate simultaneously in the figure's center, where they produce two repelling peaks. From $H$ in (\ref{H}) one infers that, between those peaks, the two lowest eigenstates are close to superpositions of $\mathopen|\uparrow\downarrow\uparrow\downarrow\nobreak\rangle$ and $\mathopen|\downarrow\uparrow\downarrow\uparrow\nobreak\rangle$, for which all the coupling energies are negative; at co-degeneracy, the superpositions become (anti)symmetric~\cite{select}. From the viewpoint of AQC, it is encouraging that the ground state of $H$ thus is globally entangled~\cite{vidal2003}; for a quantification, see~\cite{love}. With all parameters already determined from Fig.~\ref{fig3} and a few other cross-sections like it, the theoretical comparison in Fig.~\ref{fig4}b (and many other nontrivial cross-sections) can be regarded as having \emph{no} free parameters, making the agreement remarkable.

Complete agreement between experiment and theory was also obtained for a second sample, with $\Delta_1=\Delta_2=60$, $\Delta_3=130$, $\Delta_4= 110$, $J_{12}=J_{23}=300$, $J_{14}=J_{34}=330$, and $J_{13}=J_{24}=-90$ (all in mK). However, the lowest effective temperature attained was only $\sim$300~mK.

In conclusion, ferro- and antiferromagnetic couplings between four 3JJ flux qubits have been realized simultaneously with shared Josephson junctions, with a coupling strength significantly exceeding the inductive one. (Incidentally, the results also show that direct galvanic coupling can be used for more general circuits than the linear arrays considered in Ref.~\cite{butcher}.) The data fully agree with a quantum-mechanical description to the experimental accuracy. Currently, quasi-equilibrium impedance measurement already provides valuable information complementary to state readout, notably the determination of the Hamiltonian. With a faster and possibly qubit-selective detection method, the circuit used looks very promising for studying adiabatic bias manipulations.

MG was supported by Grants VEGA 1/2011/05 and APVT-51-016604, EI by the RSFQubit project, and AZ by the NSERC Discovery Grants Program. We thank A.J. Berkley, D.A. Lidar, and M.F.H. Steininger for discussions.


\begin{thebibliography}{99}
\bibitem{complete} The stability diagram is completed by connecting the opposing corners where traces corresponding to skipped crossings meet. These invisible lines correspond to the $\mathopen|\uparrow\downarrow\rangle\leftrightarrow
\mathopen|\downarrow\uparrow\rangle$ anticrossings.
\bibitem{select}  However, the minimum energy gap $E_1-E_0\approx8$~mK between $|\Psi_1\rangle$ and $|\Psi_0\rangle$ (due to fourth-order tunnelling in pertubative language) was less than the thermal energy, partly because of the small~$\Delta_2$. Therefore, we do not claim selectivity between these GHZ-type states.

\newpage
\end{thebibliography}
\end{document}